\documentclass[conference]{IEEEtran}
\IEEEoverridecommandlockouts
\usepackage{tikz} %
\usepackage{cite}
\usepackage{amsmath,amssymb,amsfonts}
\usepackage{algorithmic}
\usepackage{algorithm}
\usepackage{graphicx}
\usepackage{textcomp}
\usepackage{xcolor}
\usepackage{amsthm}
\usepackage{mathtools}
\usepackage{comment}
\usepackage[scientific-notation=true]{siunitx}
\usepackage{mathtools}

\usetikzlibrary{external}
\usepackage{pgfplots}

\def\BibTeX{{\rm B\kern-.05em{\sc i\kern-.025em b}\kern-.08em
    T\kern-.1667em\lower.7ex\hbox{E}\kern-.125emX}}

\definecolor{tab0}{RGB}{228,26,28}
\definecolor{tab1}{RGB}{55,126,184}
\definecolor{tab2}{RGB}{77,175,74}
\definecolor{tab3}{RGB}{152,78,163}
\definecolor{tab4}{RGB}{255,127,0}

\usepackage{printlen}
\uselengthunit{mm}

\newtheorem{thm}{Theorem}

\def\hammingweight{W}
\def\noise{Z}
\def\syndrome{S}
\def\intercept{c}

\def\codebook{\mathcal{C}}
\def\LLR{\text{LLR}}
\def\reliability{\Gamma}

\def\fbinary{\mathbb{F}_2}

\def\generatormatrix{G}

\begin{document}
\title{A Balanced Tree Transformation to Reduce GRAND Queries%
}

\newcommand*{\affaddr}[1]{#1} %
\newcommand*{\affmark}[1][*]{\textsuperscript{#1}}
\newcommand*{\email}[1]{\texttt{#1}}

\author{
	\IEEEauthorblockN{Lukas~Rapp\affmark[1], Jiewei~Feng\affmark[2], Muriel M\'edard\affmark[1], Ken R. Duffy\affmark[2]}\\
	\IEEEauthorblockA{\affmark[1]\textit{Massachusetts Institute of Technology} \\   
	\affmark[2]\textit{Northeastern University}\\
	\{rappl,medard\}@mit.edu, \{feng.ji,k.duffy\}@northeastern.edu}
}

\maketitle

\begin{abstract}
Guessing Random Additive Noise Decoding (GRAND) and its variants, known for their near-maximum likelihood performance, have been introduced in recent years. One such variant, Segmented GRAND, reduces decoding complexity by generating only noise patterns that meet specific constraints imposed by the linear code. In this paper, we introduce a new method to efficiently derive multiple constraints from the parity check matrix. By applying a random invertible linear transformation and reorganizing the matrix into a tree structure, we extract up to \(\log_2 n\) constraints, reducing the number of decoding queries while maintaining the structure of the original code for a code length of \(n\). We validate the method through theoretical analysis and experimental simulations.
\end{abstract}


\section{Introduction}
Guessing Random Additive Noise Decoding (GRAND) and its variants \cite{8630851,9541392,9685645,10206762,abbas2022list,yuan2025SOGRAND,10437763,10465063,9605006,9195254,10067519,galligan2023block} are universal and near Maximum Likelihood (ML) decoders that can decode any moderate
redundancy code of any structure \cite{9500279} including non-linear codes. The main idea behind GRAND algorithms is to sequentially generate putative binary noise effect sequences in decreasing order of likelihood, where the first putative noise effect that leads to a valid codeword will be the resulting decoding. 
Depending on the statistical model of the channel, the likelihood of a noise effect can be calculated accordingly and different GRAND decoders are used.

To minimize the number of queries required to identify the first valid noise sequence when decoding binary linear codes, recent works \cite{10436895,9834343,rowshan2023segmented} have made contributions by introducing segmentation techniques that utilize a specific code structure to skip invalid noise sequences. The core idea is that a candidate noise sequence $z^n=(z_1,...,z_n)\in\fbinary^n$ leads to a valid codeword only if $z^n$ satisfies certain linear constraints. Rather than generating all possible noise sequences in decreasing order of likelihood, one can focus on only generating those that meet at least one or more constraints. A detailed discussion of segmentation is provided in Section \ref{Preliminary}. The effectiveness of this approach in reducing query numbers depends on the number of constraints employed. However, to utilize multiple constraints simultaneously, the work in \cite{9834343} assumes that the subsets of bits covered by each constraint are disjoint, or equivalently, the parity check matrix of the code should have specific properties. This limits the potential query reduction for a given code, and an efficient method for extracting numerous constraints for any given code is currently lacking. In this paper, we propose a novel way to rewrite the parity check matrix of any given code to a structure that maintains the same binary linear codebook while facilitating the extraction of additional constraints for noise effect skipping. 

The remainder of this paper is organized as follows: Section \ref{Preliminary} introduces the basic setup of the decoding. Section \ref{section_tree_format} defines the Tree Structure and Balanced Tree Transformation, which rewrites the parity check matrix of any given linear code to a matrix with desirable properties. Section \ref{section_extraction} demonstrates how to leverage the Tree Structure to extract multiple constraints efficiently for decoding. Section \ref{section_simulation} provides experimental validation of the proposed approach.

\section{Preliminaries}\label{Preliminary}
Let $\generatormatrix$ be a generator matrix of a binary linear $(n,k)$. It defines the relationship $U^k\generatormatrix=X^n$ between a random message $U^k=(U_1,U_2,...,U_k)\in\{0,1\}^k$ and its coded sequence $X^n=(X_1,...,X_n) \in \{0, 1\}^n$. The set $\codebook\subset\{0,1\}^n$ records all possible $2^k$ codewords. Let $m=n-k$ and \(H \in \fbinary^{m \times n}\) be a parity check matrix of $\generatormatrix$. A codeword $X^n$ is sent through a noisy channel, and the detector receives signal $R^n=(R_1,...,R_n) \in \mathbb{R}^n$, with hard-decision demodulated signal $Y^n=X^n\oplus \noise^n$ where $\noise^n$ represents the binary noise effect and $\oplus$ represents binary addition.

Given an assumption on the distribution of noise, $f_{R|X}$ is the conditional pdf of $R_i$ conditioned on the value of $X_i$. 
The log-likelihood ratio (LLR) of a received signal $R_i$ is defined to be
\begin{align}
\LLR(R_i)=\log\frac{f_{R|X}(R_i|1)}{f_{R|X}(R_i|0)}.  \label{LLR_formula}
\end{align} 
With equation ({\ref{LLR_formula}}), the reliability $\Gamma_i$ of the received signal $R_i$ is defined to be $\Gamma_i=|\LLR(R_i)|$.

It follows from \cite{9414615} that the probability of the noise $z^n$ given the received signal $R^n$ is 
\begin{align}
P({\noise}^n=z^n|R^n) \propto e^{-\sum_{i:z_i=1}\reliability_i},   \label{noise_likelihood}
\end{align}
which says that, regardless of the distribution of noise, it suffices to find the rank order of the values $\sum_{i:z_i=1}\reliability_i$ from smallest to highest to rank order the probability of noise patterns. This idea is one of the essential foundations of GRAND and its variants, resulting in an efficient generation of noise patterns in decreasing order of probability. For each generated noise effect $z^n$, GRAND tests whether the inversion of $z^n$ from the demodulated sequence leads to a codeword, i.e., if $z^n\oplus Y^n\in\codebook$. This is equivalent of checking if $H(z^n\oplus Y^n)^T=\textbf{0}$ where $\textbf{0}$ represents a zero vector. The first time the decoder finds a noise sequence $z^n$ that satisfies $H(z^n\oplus Y^n)^T=\textbf{0}$ it will output the codeword $z^n\oplus Y^n$ as the decoded sequence. This implies that a noise sequence that leads to a valid codeword must satisfy some constraints indicated by the formula $H(z^n)^T=H(Y^n)^T$. Usually the term $H(Y^n)^T$ is denoted as $\syndrome^n=(\syndrome_1,...,\syndrome_{m})^T$ and is called the syndrome.

We now give an outline of Segmented GRAND \cite{rowshan2023segmented}, which uses the parity check matrix to skip a subset of the noise effects. Let $H_{i,:}$ denote the $i$-th row of $H$. Then, taking the first row $H_{1,:}$ of $H$ indicates that a valid noise effect must satisfy $H_{1,:} (z^n)^T= \syndrome_1$ imposing a constraint on the number of ones in specified coordinates of $z^n$. 
Therefore, only noise effects that meet this first constraint need to be generated; otherwise, they will not pass the parity check. 
Furthermore, the ones in $H_{1,:}$ specify which bits are covered by the first constraint. Therefore, $z^n$ can be split into two sub-sequences $z^n_{(1)},z^n_{(2)}$ where $z^n_{(1)}$ contains the bits that are covered by the first constraint and $z^n_{(2)}$ contains the remaining bits of $z^n$. This allows the sub-sequences to be generated separately, where a decoder only generate sub-sequences $z^n_{(1)}$ that satisfy the first constraint and merge them with sub-sequences $z^n_{(2)}$ to produce $z^n$. If the second row $H_{2,:}$ is also considered, which specifies another constraint, then the decoder can only generate the noise effects that satisfy both constraints. It is proposed in \cite{9834343,rowshan2023segmented} that if the first and second constraints cover different bits in $z^n$, then $z^n$ can be separated into three sub-sequences $z^n_{(1)},z^n_{(2)},z^n_{(3)}$ instead, where $z^n_{(1)}$ is covered by the first constraint and $z^n_{(2)}$ is covered by the second constraint. The noise effect generator can then generate noise effects $z^n$ by only generating the sub-sequences that satisfy their respective constraints. The idea is then to use multiple constraints, where each constraint should cover disjoint sets of bits in $z^n$ in order to be used in the noise effect generator. 
A generalization of this approach is introduced in \cite{9834343}, which requires that the $i$-th constraint cover a subset of bits covered by the $(i-1)$-th constraint. The requirement allows the extracted constraints to be rewritten into constraints covering disjoint sets.
\section{Tree Structure and balanced tree transformation}\label{section_tree_format}
As noted earlier, Segmented GRAND relies on extracting constraints from the parity check matrix that apply to disjoint sets of transmitted bits. The more constraints we can incorporate into noise generation, the greater the reduction in the number of decoding queries. Therefore, efficiently extracting these constraints is of key interest. We first demonstrate that any parity check matrix can be transformed into a special structure, which we call the \textit{Tree Structure} that results in an equivalent code, i.e., the codebook is the same except for a fixed permutation of the codewords \cite[Def. 3.5]{moonErrorCorrectionCoding2005}.
Theorem \ref{thm_uniform_matrix} proves that with a suitable invertible matrix identified using a stochastic approach, a given parity check matrix can be rewritten into an equivalent parity check matrix with advantageous properties. In section \ref{section_extraction}, we outline how the Tree Structure facilitates the extraction of disjoint constraints from the parity check matrix.

\begin{figure}
    \centering
    \includegraphics[width=0.8\linewidth]{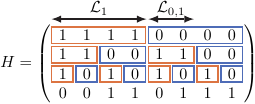}
    \caption{Example of Tree Structure. For each row, adjacent entries with $1$ are highlighted in orange, while adjacent entries with $0$ are highlighted in blue. Each rectangle represents a subset of the columns, for example, the orange rectangle in the first row represents $\mathcal{L}_1$ and the second rectangle in the second row represents $\mathcal{L}_{0,1}$. As demonstrated in the figure, the columns are divided into two disjoint sets based on the first row, and then each of the subsets is further divided into subsets by the second row and the third row as well. This exhibits a branching Tree Structure and hence we call this the Tree Structure. Note that as demonstrated in the figure, the number of subsets that the bits can be divided into is bounded above by $2^m$ and $\log_2n$, where $m$ is the number of rows of $H$ and $n$ is the number of columns of $H$.} %
    \label{fig:tree_example}
    \vspace{-10pt}
\end{figure}

For a given parity check matrix, we can rearrange its columns so that the binary values represented by the columns form a non-decreasing sequence. This reordering can be applied to any parity check matrix through a column permutation and results in an equivalent code. For example, consider the following matrix $H$ and its Tree Structure matrix $H'$:
\[ H=\left( \begin{array}{cccc}
1 & 0 & 0 & 1 \\
0 & 1 & 1 & 1 \\
0 & 1 & 0 & 1
\end{array} \right)
\rightarrow
H'=\left( \begin{array}{cccc}
1 & 1 & 0 & 0 \\
1 & 0 & 1 & 1 \\
1 & 0 & 1 & 0
\end{array} \right)
\]
where the integer value for each column in $H'$ is $7,4,3,2$. Note that the same column permutation must also be applied to the generator matrix $\generatormatrix$ to obtain the corresponding generator matrix $\generatormatrix'$ for $H'$. Hence, without loss of generality, we can assume that the integer values for the columns in the parity check matrix $H$ form a non-decreasing sequence. This column permutation gives the transformed parity check matrix a Tree Structure in the following sense \cite{rappErrorandErasureDecodingProduct2022}: Let $\ell\in\mathbb{N}$ and $v_1,v_2,...,v_\ell\in\fbinary$, and $H_{i,j}$ denote the entry of $H$ at $i$-th row and $j$-th column. Define
\begin{align}
    \mathcal{L}_{v_1, \dots, v_\ell} \coloneqq \{
        j | H_{1, j} = v_1, \dots, H_{\ell, j} = v_\ell.
    \}\label{def_leaf_set}
\end{align}
which denotes the subset of columns where the top $\ell$ entries are $v_1,v_2,...,v_\ell$. The following properties hold for $\ell=1,...,m$ given $H$: 1), $\mathcal{L}_{v_1, \dots, v_\ell}=\mathcal{L}_{v_1, \dots, v_{\ell},1} \cup \mathcal{L}_{v_1, \dots, v_{\ell},0}$, meaning the columns are divided into smaller subsets as more rows in $H$ are considered; 2) The elements in $\mathcal{L}_{v_1, \dots, v_\ell}$, if they exist, are consecutive integers and, 3) $\cup_{v_1,v_2,...,v_\ell}\mathcal{L}_{v_1,v_2,...,v_\ell}=\{1,...,n\}$ for all $\ell\leq m$, meaning that the disjoint subsets resulting from considering the first $\ell$ rows cover all the columns of $H$. An example is given in Fig. \ref{fig:tree_example}.

The division of columns similarly partitions any noise effect $\noise^n=(\noise_1,\noise_2,...,\noise_n)=X^n\oplus Y^n$ into corresponding sub-sequences, denoted as $\noise^n(\mathcal{L}_{v_1,...,v_\ell})$
\begin{align*}
    :=(Z_{\min(\mathcal{L}_{v_1,...,v_\ell})},Z_{\min(\mathcal{L}_{v_1,...,v_\ell})+1},...,Z_{\max(\mathcal{L}_{v_1,...,v_\ell})}).
\end{align*}
 This represents the sub-sequence of $Z^n$ with entries in the columns belonging to $\mathcal{L}_{v_1,...,v_\ell}$. Using the example in Fig.\ref{fig:tree_example}, $\noise^n(\mathcal{L}_{1,1})=(\noise_1,\noise_2)$, $\noise^n(\mathcal{L}_{1,0})=(\noise_3,\noise_4)$, $\noise^n(\mathcal{L}_{0,1})=(\noise_5,\noise_6)$ and $\noise^n(\mathcal{L}_{0,0})=(\noise_7,\noise_8)$. Note that by the third property of (\ref{def_leaf_set}), the division of $\noise^n$ into the sub-sequences can be done reversely to recover $\noise^n$ using all the sub-sequences. We will call the sub-sequences $\noise^n(\mathcal{L}_{v_1,...,v_\ell})$ the \textbf{\textit{sub-effects}}.

Next, we introduce the Balanced Tree Transformation (BTT), which rewrites a parity check matrix using a random invertible matrix so that each of the rows in the resulted matrix is independent and, with probability bounded below by a constant, around half of the entries have value one. This property provides a suitable condition for applying the framework proposed in Segmented GRAND \cite{rowshan2023segmented} after extracting constraints. We first explain why this property is desired, followed by Theorem \ref{thm_uniform_matrix} explaining how the property is achieved. 

As pointed out by \cite{ahsanullah2013introduction}\footnote{See Example 10.3 in \cite{ahsanullah2013introduction} for more detail.}, the $q$-th sample quantile of random variables $\reliability_1,\reliability_2,..,\reliability_n$ converges to $F^{-1}(q)$ where $F$ is the cumulative distribution function (CDF) of $\reliability_i$ and $F^{-1}(s)=\inf\{u:F(u)\geq s\}$. As a result, we can expect that the plot of rank-ordered reliabilities versus the rank is approximately the function $F^{-1}(i/n)$ for $i\in[1,n]$. When the rank-ordered reliability plot is approximately a straight line \cite{duffy2022_ordered}, the reliabilities can be approximated by 
\begin{align}
    &\reliability_i\approx \beta \pi^n(i)+\intercept,
  &\beta\approx \frac{1}{n}\cdot\frac{dF^{-1}(s)}{ds}=:\frac{(F^{-1})'}{n} \label{reliability_approximation}  
\end{align}
for some $\beta,\intercept\in\mathbb{R}$ and $\pi^n(i)$ is the rank of the reliability of the $i$-th bit within the $n$ received bits. 
In \cite{rowshan2023segmented}, the $n$ bits are separated into different groups called segments. Let $\pi^n_{*}(i)$ denote the rank of $\reliability_i$ within the segment that contains the $i$-th bit. It is implicitly assumed in \cite{rowshan2023segmented} that 
\begin{align}
 \left\{\pi^n_{*}(i)=\pi^n_{*}(j) \right\}\implies\left\{\reliability_i\approx\reliability_j\right\},\label{implication_assumption}
\end{align} 
which says that, by (\ref{noise_likelihood}), an error in the $i$-th bit has the same impact to the order of the likelihood of noise effects as the $j$-th bit if their reliabilities have the same rank within each of their segments. 
This further assumes that the size of each segment is equal by the following idea: Let $n_i$ denote the number of bits that belong to the segment which contains the $i$-th bit, then applying the same argument for (\ref{reliability_approximation}) to each segment by noting that the reliabilities of bits in each segment is independent between segments, we have 
\begin{align}
    \reliability_i\approx \frac{(F^{-1})'}{n_i}\pi^n_*+c. \label{reliability_appro_segment}
\end{align}
By (\ref{reliability_appro_segment}), it is desired that $n_i=n_j$ for all $i\neq j$ in order to satisfy (\ref{implication_assumption}). With BTT, a given parity check matrix can be rewritten in a Balanced Tree Structure where the sets $\mathcal{L}(v_1,...,v_\ell)$ for fixed $\ell$ can be seen as the segments. The size of each $\mathcal{L}(v_1,...,v_\ell)$ is approximately the same, and hence BTT provides a suitable condition of the segments to be used in Segmented GRAND.

The idea of BTT is to apply a random invertible matrix to $H$ before rewriting it in Tree Structure and the probability that we get a transformed matrix with the desired property is bounded below by $0.22$ as $m\rightarrow \infty$. To provide a theoretical validation, Theorem \ref{thm_uniform_matrix} is first proved without invertible assumption on the random transformation. After Theorem \ref{thm_uniform_matrix}, we will restrict to invertible random transformation so that the rewritten code is equivalent to the given code.
\begin{thm}\label{thm_uniform_matrix}
    Let $H^{m,\kappa m}$ be a given $m\times \kappa m$ binary matrix, for $\kappa\in \mathbb{N}$. Separate $H^{m,\kappa m}$ into $\kappa$ non-overlapping sub-matrices with dimension $m \times m$ and denote each of the sub-matrices as $H^{m,\kappa m}_{(j)}$. Let $\mathcal{H}^{m,\kappa m}:= \{H^{m,\kappa m}|H^{m,\kappa m}_{(j)} \text{ is invertible for all }j\leq \kappa \}$ be the set of parity check matrices whose sub-matrices are all invertible. Let $\widetilde{H}^{m,\kappa m}=A^{m,m}H^{m,\kappa m}$, where $A^{m,m}\in\fbinary^{m\times m}$ is a random binary matrix whose entries are i.i.d Bernoulli distributed with rate $0.5$. Then: 1), each $\widetilde{H}^{m,\kappa m}_{(j)}=A^{m,m}H^{m,\kappa m}_{(j)}$ is uniformly distributed over all binary $m\times  m$ matrices given $H^{m,\kappa m}\in\mathcal{H}^{m,\kappa m}$ and hence can be treated as matrix with i.i.d entries from Bernoulli distribution with rate $0.5$. 2), the rows in $\widetilde{H}^{m,\kappa m}$ are independent to each other. 3), given $\delta>0$, let $\{H^{m,\kappa m}\}_{m\in\mathbb{N}}$ be any given sequence of parity check matrices such that $H^{m,\kappa m}\in\mathcal{H}^{m,\kappa m}$ for all $m$, and let $H^{m,\kappa m}_{{(j)},i,s}$ denote the entry of $H^{m,\kappa m}_{(j)}$ at the $i$-th row and $s-th$ column, then 
    \begin{align}
        \lim_{m\rightarrow \infty}P\left(\bigcup_{j=1}^{\kappa}\bigcup_{i=1}^m\Bigg\{\left|\sum_{s=1:m}\frac{\widetilde{H}^{m,\kappa m}_{(j),i,s}}{m}-0.5\right|>\delta\Bigg\}\right)=0.
    \end{align}
\end{thm}

\begin{proof}
    Let $\widetilde{h}\in\fbinary^{m\times m}$ be any matrix,
    \begin{align*}
       &P\left(\widetilde{H}^{m,\kappa m}_{(j)}=\widetilde{h}|H^{m,\kappa m}_{(j)}\right)\\
       =&P\left(A^{m,m}=\widetilde{h}[H^{m,\kappa m}_{(j)}]^{-1} \middle|H^{m,\kappa m}_{(j)}\right)=2^{-m^2}.
    \end{align*}
    Hence $\widetilde{H}^{m,\kappa m}_{(j)}=A^{m, m}H^{m,\kappa m}_{(j)}$ is uniformly distributed over all binary $m\times  m$ binary matrices. With coupling technique \cite{Thorisson2000CouplingSA}, $\widetilde{H}^{m,\kappa m}_{(j)}$ can be treated as a matrix with i.i.d entries from Bernoulli distribution with rate $0.5$. This completes the first claim.

    Notice that $\widetilde{H}^{m,\kappa m}_{i,j}=\sum_{s=1}^m A^{m,\kappa m}_{i,s}H^{m,\kappa m}_{s,j} \mod 2$ where $H^{m,\kappa m}$ is given and the rows of $A^{m,\kappa m}$ are independent. This completes the second claim.

    To prove the third claim, putting the first claim, Hoeffding's inequality and union bound together yields that, given $\delta>0$, 
    
    \begin{align*}
      \lim_{m\rightarrow \infty}P\left(\bigcup_{i=1}^m\Bigg\{\left|\sum_{s=1:m}\frac{\widetilde{H}^{m,\kappa m}_{(j),i,s}}{m}-0.5\right|>\delta\Bigg\}\right)=0,
    \end{align*}
    which means we can expect that a sub-matrix has almost half the number of ones and zeros within each row. By the union bound, we are able to conclude that this happens to all sub-matrices of $\widetilde{H}^{m,\kappa m}$ simultaneously as we have a fixed code rate $1/\kappa$, which finishes the third claim. 
\end{proof}
The first result of Theorem \ref{thm_uniform_matrix} states that a parity check matrix, whose sub-matrices are invertible, can be transformed into a random matrix whose entries in each sub-matrix can be treated as i.i.d. variables with equal probability of being one or zero. This further leads to the properties stated in the second and third results. Choosing $\delta$ close to zero and applying the third result shows that as $m\rightarrow\infty$, we can expect that the number of ones is approximately the same as the number of zeros in each row of $\widetilde{H}^{m,\kappa m}$. Together with the fact that each row is independent in $\widetilde{H}^{m,\kappa m}$, we can expect that the size of each $\mathcal{L}(v_1,...,v_\ell)$ of $\widetilde{H}^{m,\kappa m}$ in Tree Structure is approximately the same, that is, the set of columns are divided into subsets with approximately equal sizes. The name Balanced Tree Transformation is given based on this property.

To retain the same structure of the code represented by $H^{m,\kappa m}$, we further require that the random transformation $A^{m,m}$ is invertible. The probability that a random binary matrix $A^{m, m}$ is invertible is bounded below by 0.22 as $m\rightarrow \infty$ \cite{BRENNAN1987311,waterhouse1987often,kolchin1999random}. Therefore, by the union bound, the probability that $A^{m,m}$ is invertible and the size of each $\mathcal{L}(v_1,...,v_\ell)$ of $\widetilde{H}^{m,\kappa m}$ in Tree Structure is approximately the same is bounded below as $m\rightarrow \infty$. This provides an asymptotic reasoning that we can transform parity check matrices into Tree Structure with desired properties with positive probability bounded below, which is desired in the framework proposed by \cite{rowshan2023segmented}. We call the procedure of multiplying $H$ by a random matrix the \textit{Balance Transformation}. Additionally, the procedure of Balance Transformation followed by rewriting the resulted matrix in Tree Structure is called \textit{Balanced Tree Transformation}. 

As BTT is invertible and hence does not change the code, we only need to apply the transformation once for any given parity check matrix and use the rewritten parity check matrix and generator matrix for application. Therefore, this transformation will not add additional complexity to the decoding.

\section{Application of Tree Structure}\label{section_extraction}
Next, we explain how we can extract constraints from $H$ in Tree Structure. Recall that a generated noise effect $z^n$ will lead to a valid codeword if and only if $H(z^n)^T=\syndrome^m=H(Y^n)^T$ where each row of $H$ indicates a constraint that any valid noise effect must satisfy. Let $\text{SUPP}(z^n):=\{i|z_i=1\}$ denote the support of $z^n$ which is the set of entries with value one. Then the constraint defined by the $\ell$-th row of $H$ can be expressed as $H_{\ell,:}(z^n)^T=\syndrome_\ell$ which is equivalent to 
\begin{align}
\left|\text{SUPP}(H_{\ell,:})\cap\text{SUPP}(z^n)\right|\equiv \syndrome_\ell \mod 2.\label{syndrom_constraint}
\end{align}
Equation (\ref{syndrom_constraint}) implies that the $\ell$-th row of $H$ and the corresponding syndrome $\syndrome_\ell$ specifies the parity of the number of ones in a subset of any valid noise effect. With the example given in Fig. \ref{fig:tree_example}, $H_{2,:}$ specifies that the parity of the number of ones in the first, second, fifth, and sixth bits is equal to $\syndrome_2$: If $\syndrome_2=1$, then there should be an odd number of ones in the first, second, fifth and sixth bits in any valid noise effect. If, on the contrary, $\syndrome_2=0$, then there should be an even number of ones in the first, second, fifth, and sixth bits instead. We can also notice that in Fig. \ref{fig:tree_example}, the two constraints imposed by the first two rows both cover the first two bits, which fails to meet the assumption in \cite{9834343} that the constraints cover disjoint subsets of bits.
We now introduce how we can use the Tree Structure to avoid generating invalid noise patterns with overlapping constraints.

Let $\hammingweight(z^n)=\sum_{i=1}^nz_i = |\text{SUPP}(z^n)|$ denote the Hamming Weight of a noise effect $z^n$ which counts how many bits are one. 
Rewriting (\ref{syndrom_constraint}) in terms of Hamming Weight leads to
\begin{align}
\sum_{v_1,v_2,...,v_{\ell-1} \in \{0, 1\}}\hammingweight(\noise^n(\mathcal{L}_{v_1,...,v_{\ell-1},1}))\equiv \syndrome_\ell \mod 2.\label{syndrom_contrains_1}
\end{align}
For simplicity of notation, denote $w^{v_1v_2...v_\ell}=\hammingweight(\noise^n(\mathcal{L}_{v_1,...,v_{\ell}}))$ as the Hamming Weight of the sub-effect $Z^n(\mathcal{L}_{v_1,v_2,...,v_\ell})$, then with the first property of (\ref{def_leaf_set}) and the first $\ell$ constraints expressed in the form of (\ref{syndrom_contrains_1}), we get the following linear equations in the binary field:
\begin{equation}
\begin{aligned}
    \sum_{v_2,...,v_{\ell} \in \{0, 1\}} w^{1v_2...v_\ell}&\equiv \syndrome_1 \mod 2 \\
    \sum_{v_1,v_3,v_4,...,v_{\ell} \in \{0, 1\}} w^{v_11v_3...v_\ell}&\equiv \syndrome_2 \mod 2 \\
    &\vdots\\
    \sum_{v_1,...,v_{\ell-1} \in \{0, 1\}} w^{v_1...v_{\ell-1}1}&\equiv \syndrome_\ell \mod 2 
\end{aligned}
\label{linear_equations_general}
\end{equation}
where there are $2^\ell$ variables on the left, and we have $\ell$ equations, hence we can rewrite the equations to express $\ell$ variables in terms of the others. While the choice of the $\ell$ variables is not unique, here we choose the variables of the form $w^{v_1...v_\ell}$ with $\sum_{i=1}^\ell v_i=1$ to be the $\ell$ variables of interest, which corresponds to the set of sub-effects $\{Z^n(\mathcal{L}_{v_1,v_2,...,v_\ell})|\sum_{i=1}^\ell v_i=1 \}$ called the set of \textit{chosen sub-effects}. The generation of noise sequences $z^n$ can be guided by (\ref{linear_equations_general}), i.e., we only need to generate putative noise effects $z^n$ whose parities of its sub-sequences satisfy (\ref{linear_equations_general}). Using the example in Fig.\ref{fig:tree_example}, the linear equations corresponding to the first three rows of $H$ are
\begin{align*}
w^{100} \equiv w^{111}+w^{110}+w^{101}+\syndrome_1 \mod 2\\
w^{010} \equiv w^{111}+w^{110}+w^{011}+\syndrome_2 \mod 2\\
w^{001} \equiv w^{111}+w^{110}+w^{011}+\syndrome_3 \mod 2
\end{align*}
where the chosen sub-effects are $Z^n(\mathcal{L}_{1,0,0})$, $Z^n(\mathcal{L}_{0,1,0})$ and $Z^n(\mathcal{L}_{0,0,1})$. 

To generate a noise effect $z^n$ using the linear equation (\ref{linear_equations_general}), one first generates its non-chosen sub-effects. 
Secondly, the parities of the Hamming Weights of the chosen sub-effects are determined from (\ref{linear_equations_general}) by inserting the parities of the non-chosen sub-effects. Thirdly, each chosen sub-effect is generated based on the parity of its Hamming Weight obtained from (\ref{linear_equations_general}).
Finally, combine all the sub-effects to recover noise effect $z^n$. This approach efficiently avoids generating a subset of the noise effects that will not lead to a valid codeword and hence reduces the number of queries before reaching the first valid noise effect that corresponds to a codeword.

Note that if the set $\mathcal{L}_{v_1,...,v_j}$ is empty, then there will be no $w^{v_1,...,v_j}$ in the linear equation. This implies that, for the $\ell$ variables that are chosen to be put on the left of the linear equations (\ref{linear_equations_general}), all the corresponding sets in $\{ \mathcal{L}_{v_1,...,v_\ell}|\sum_{i=1}^\ell v_\ell=1\}$ can not be empty. Hence, to be able to extract $\ell$ constraints using the first $\ell$ rows of $H$ in Tree Structure, we should ensure that the sets in $\{ \mathcal{L}_{v_1,...,v_\ell}|\sum_{i=1}^\ell v_\ell=1\}$
are nonempty. For future reference, the elements in $\{ \mathcal{L}_{v_1,...,v_\ell}|\sum_{i=1}^\ell v_\ell=1\}$ are called the \textit{chosen column subsets}. Equivalently speaking, to extract as many constraints as possible, we should maximize $\ell$ while ensuring that the chosen column subsets are nonempty.
While many well-designed codes automatically satisfy this assumption, we can also use BTT to achieve this goal due to the third result of Theorem \ref{thm_uniform_matrix}.

\section{Simulation Results}\label{section_simulation}
\begin{figure}
    \centering
    \includegraphics{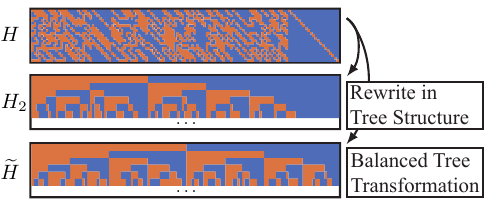}
    \vspace{-3pt}
    \caption{Example of rewriting BCH(127, 106) code.}%
    \label{fig:enter-label}
    \vspace{-13pt}
\end{figure}

In this section, we will provide experimental verification of BTT in addition to the theoretical results by Theorem \ref{thm_uniform_matrix}. The simulation will be conducted using a decoding framework based on \cite{rowshan2023segmented}. Fig. \ref{fig:enter-label} demonstrates the resulting parity check matrix by applying the procedures introduced in Section \ref{section_tree_format} to a Bose–Chaudhuri–Hocquenghem (BCH) code. 
Three figures are presented in Fig. \ref{fig:enter-label}: The top figure represents the parity check matrix $H$ for BCH(127,106) code, where orange pixels represent entries with a value of one and blue pixels represent entries with a value of zero. The middle figure represents the first six rows of the parity check matrix $H_2$, which results from rewriting $H$ in Tree Structure. 
The bottom figure represents the first six rows of the parity check matrix $\widetilde{H}$, obtained by first multiplying $H$ by a random invertible matrix $A^{11,11}$ followed by rewriting $A^{11,11}H$ into Tree Structure. 
As depicted in Fig.~\ref{fig:enter-label}, at least six rows in the rewritten parity check matrix can be used in the sense of (\ref{linear_equations_general}), and each of the first four rows is split into blocks with approximately equal sizes. This indicates the effectiveness of BTT for easy extraction of constraints.

\begin{figure}
    \centering
    \includegraphics{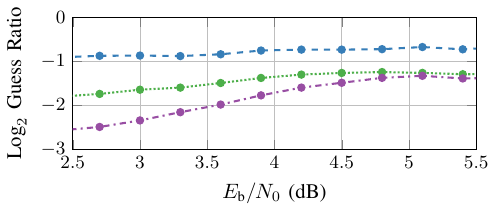}
    \includegraphics{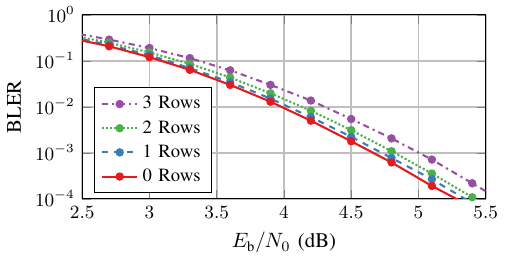}
    \vspace{-3pt}
    \caption{Example for query number reduction using the Balanced Tree Transformation on BCH (127,106) code.} %
    \label{fig:BCH_127_113_nguess}
    \vspace{-13pt}
\end{figure}

Next, Fig.~\ref{fig:BCH_127_113_nguess} demonstrates the reduction in the number of noise effect queries achieved by extracting more constraints using multiple rows from the parity check matrix. The top plot demonstrates the reduction in query number when using different numbers of rows in the parity check matrix $H$. We use $\ell$ to denote the number of rows in $H$ that are used to extract constraints and skip noise effects. The blue line represents the ratio in $\log_2$ scale between $\ell=1$ and $\ell=0$ where the latter is the original ORBGRAND with no noise effect skipped. Similarly, the green and purple lines represent the complexity reduction of $\ell=2$ and $\ell=3$ compared to \(\ell=0\), respectively. 
As depicted by the top plot, each additional row used to skip noise patterns reduces the query number by up to a factor of two at lower $E_\text{b} / N_0$.
The bottom plot shows the BLER for different $\ell$. Segmented GRAND experiences a slight performance loss compared to the original ORBGRAND algorithm in exchange for reduced query complexity. To be precise, \num[scientific-notation=false]{0.06}, \num[scientific-notation=false]{0.16}, and \SI[scientific-notation=false]{0.33}{\dB} loss in SNR at a target BLER of \num[scientific-notation=false]{e-3} for \(l = 1, 2, 3\), respectively.
This is caused by partitioning the noise effects into sub-effects where the latter is generated separately. The figure highlights the significance of the BTT, as it enables a systematic framework for efficiently extracting multiple constraints, leading to a substantial reduction in query complexity.

\section{Conclusion}
In this work, we introduce a method called Balanced Tree Transformation for rewriting the parity check matrix of a linear code, along with a corresponding query-skipping approach that leverages the transformed matrix. Theorem \ref{thm_uniform_matrix} provides theoretical validation of the approach whose assumptions can be further generalized using properties of random matrices. The BTT restructures the parity check matrix to allow for the systematic extraction of multiple noise effect constraints, resulting in an equivalent code. This enables GRAND decoders to operate with increased speed and efficiency.

\section{Acknowledgment}
This work was supported by the Defense Advanced Research Projects Agency (DARPA) under Grant HR00112120008.

\bibliography{main.bbl}

\end{document}